\documentclass[reprint,superscriptaddress,twocolumn,aps,pre,showpacs,floatfix]{revtex4}
\usepackage{graphicx,float}
\usepackage{amsbsy,amssymb,amsmath,bm,ulem,enumerate}  
\DeclareMathOperator{\arccot}{arccot}
\usepackage{color}

\newcommand{\g}{eps}

\usepackage{ulem} 
\newcommand{\removed}[1]{}

\renewcommand{\removed}[1]{\sout{#1}}

\newsavebox{\mysquare}
\savebox{\mysquare}{\textcolor{black}{\rule{1mm}{3mm}}}

\begin{document}
\title{Information flow and optimal protocol 
for Maxwell's demon single electron pump}
\author{J. Bergli}
\affiliation{Department of Physics, University of Oslo, P.O.Box 1048
  Blindern, N-0316 Oslo, Norway}
\author{Y. M. Galperin}
\affiliation{Department of Physics, University of Oslo, P.O.Box 1048
  Blindern, N-0316 Oslo, Norway}
\affiliation{A. F. Ioffe Physico-Technical Institute RAS, 194021 St.
Petersburg, Russian Federation}
\author{\framebox{N. B. Kopnin}}
\affiliation{Low Temperature Laboratory, Aalto University, P.O. Box 15100,
FI-00076 Aalto, Finland}
\affiliation{L.D.\ Landau Institute for Theoretical Physics, 117940 Moscow, Russia}

\begin{abstract}
We study the entropy and information flow in a 
Maxwell demon device based on a single-electron transistor with controlled gate
potentials.  
We construct the protocols for measuring the charge states and manipulating the gate
voltages which minimizes irreversibility for (i)~constant input power
from the environment or (ii)~given energy gain. Charge measurement is modeled
by a series of detector readouts for time-dependent gate potentials, and 
the amount of information obtained is determined. The protocols optimize irreversibility 
that arises due to (i)~enlargement of the configuration space on opening the barriers, 
and (ii)~finite rate of operation. These optimal protocols 
are general and apply to all systems where barriers between different regions 
can be manipulated. 
\end{abstract}

\maketitle

\section{Introduction}

Thermodynamic properties of driven small systems, where fluctuations
play an important role, are a focus of interest,
see~\cite{Seifert2012,Sagawa2012} for reviews.  It has been shown that
some mesoscopic systems can implement the so-called Maxwell's demon
(MD) process~\cite{Maxwell_Demon2003} in which the information of an
observer is used to convert the energy of thermal fluctuations into
free energy without performing work on the
system~\cite{Maruyama2009,Toyabe2010,Averin2011a, schaller2011}. This
process would violate the second law of thermodynamics if the
information about the system could be obtained and deleted without
expenditure of work or dissipation of heat.  Landauer's
principle~\cite{Sagawa2012,Landauer1961}, which equates the erasure of
information to generation of heat, restores the second law.  Among
various mesoscopic systems, single-electron tunneling (SET) devices
are particularly promising for studies of nanoscale thermodynamics.
They manipulate individual electrons in systems of metallic tunnel
junctions~\cite{Averin1986}. Being efficiently controlled in
experiment, they allow one to design and implement various
architectures.  Recently it was demonstrated that a single-electron
pump, monitored by a charge detector able to resolve individual
electrons, can be adapted to act as MD~\cite{averin2011}.  To act
efficiently, the detection should be very fast and error-free. This
can be achieved by lowering the electron tunneling rates in the pump,
in particular, by using hybrid normal-metal--insulator--superconductor
junctions~\cite{Maisi2011,Saira2012}.

Here we analyze thermodynamics and information flow in a model device
close to that proposed
in~\cite{averin2011,schaller2011,Averin2011a,Esposito2010,Strasberg2013}:~a
SET monitored by a tunnel 
junction
 with closed-loop feedback
manipulating the time-dependent tunneling rates across its junctions.
We first discuss the amount of information obtained on the system, and
thereby the amount of heat necessarily dissipated in the erasure
process.  The analysis shows that the process is irreversible for two
reasons: (i)~irreversible expansion of the available configuration
space at the opening of a barrier or after a measurement, and (ii)~the
finite rate of operation. An irreversible process means that some free
energy is converted into heat instead of mechanical work.  We propose
the way to minimize this lost work by implementing a protocol that
uses reversible expansion of configuration space similar to that
suggested in Ref.~\cite{horowitz2011}.  The main idea is to enlarge
the configuration space in a situation where in equilibrium the
probability of occupying the extra configuration space is small. With
this protocol, the lost work could be reduced to zero at infinite
operation time. However, the power extracted from the device under
infinitely slow operation implied in Ref.~\cite{horowitz2011} would
also vanish and make the entire machine meaningless. Here we construct
an optimal finite-time protocol for both the measurement process and
the gate manipulation. We consider the detection process as a series
of successive charge measurements separated with time interval $\tau$
(see, e.g., ch.~7 of review~\cite{Seifert2012}). Our idea is to
identify $\tau$ as the response time of the detector and to relate the
manipulation and measurement processes in an optimal way that
minimizes the lost work under two different conditions of
(i)~\textit{given heat flux into the system} or (ii)~\textit{
  given energy gain}.  The paper is organized as follows: In
Sec. \ref{sec:model} we present the model for the MD device. The
thermodynamics and information flow in this model is discussed in
Sec. \ref{sec:info} and the optimal protocol which minimizes the
entropy production in Sec. \ref{sec:optimal}. A short discussion of
the results is given in Sec. \ref{sec:disc}.

\section{Model}\label{sec:model}

Consider a single particle, which can be in one of the states:
in the initial $E_0$ or the final state $E_{M}$ on the right or left
lead, respectively, or on one of the intermediate islands with
energies $E_i$, $i=1,\ldots, M-1$, where $E_{M}> E_{M-1}> \ldots >
E_1>E_0$. The islands are separated from each other and from the leads
by potential barriers. Manipulating with these barriers one can open
or close them. Closed barrier implies zero transparency, while open
barrier still has a finite transmission probability to ensure that the
states on the leads and on the island are well defined.
The barriers are treated
as sliding doors, so that they can move without
requiring work. In other words, we focus on the thermodynamic 
processes in the device itself, without taking into account the energy
dissipation in the control unit and measurement device used for detecting a
particle on an intermediate island.
The system is in contact with a thermal reservoir at temperature $T$ (we measure temperature in energy units, $k_B=1$).  
The model can be, in principle, implemented in a
normal-metal--insulator--superconductor (NIS)
single-electron pump such as discussed in~\cite{averin2011}.

We study the process when the system passes through the following
states:
\begin{enumerate}[(1)]
\item  
Initially the particle is in the right lead and the gate
between the right lead and the neighboring island is open\\
\centerline{
\begin{picture}(60,10)(0,10)
\put(47,0){\line(0,1){10}}
\put(47,0){\line(1,0){10}}
\put(47,10){\line(1,0){10}}
\put(34,5){\circle{10}}
\put(12,5){\circle{10}}
\put(3,5){\circle*{1}}
\put(0,5){\circle*{1}}
\put(22,0){\usebox{\mysquare}}
\put(42,10){\usebox{\mysquare}}
\put(52,5){\circle*{5}}
\end{picture}
}\\[-0.1in]
\item 
The
particle jumps to the island and is detected to be there by the
measurement device\\
\centerline{
\begin{picture}(60,10)(0,10)
\put(47,0){\line(0,1){10}}
\put(47,0){\line(1,0){10}}
\put(47,10){\line(1,0){10}}
\put(34,5){\circle{10}}
\put(12,5){\circle{10}}
\put(3,5){\circle*{1}}
\put(0,5){\circle*{1}}
\put(22,0){\usebox{\mysquare}}
\put(42,10){\usebox{\mysquare}}
\put(34,5){\circle*{5}}
\end{picture}
}\\[-0.1in]
\item On detection we switch the gates
so that the particle cannot jump back to the right lead
\\
\centerline{
\begin{picture}(60,10)(0,10)
\put(47,0){\line(0,1){10}}
\put(47,0){\line(1,0){10}}
\put(47,10){\line(1,0){10}}
\put(34,5){\circle{10}}
\put(12,5){\circle{10}}
\put(3,5){\circle*{1}}
\put(0,5){\circle*{1}}
\put(22,10){\usebox{\mysquare}}
\put(42,0){\usebox{\mysquare}}
\put(34,5){\circle*{5}}
\end{picture}
}\\[-0.1in]
\item
The
particle jumps from the first island to the second island and our
measuring device registers that the first island is empty
\\
\centerline{
\begin{picture}(60,10)(0,10)
\put(47,0){\line(0,1){10}}
\put(47,0){\line(1,0){10}}
\put(47,10){\line(1,0){10}}
\put(34,5){\circle{10}}
\put(12,5){\circle{10}}
\put(3,5){\circle*{1}}
\put(0,5){\circle*{1}}
\put(22,10){\usebox{\mysquare}}
\put(42,0){\usebox{\mysquare}}
\put(12,5){\circle*{5}}
\end{picture}
}\\[-0.1in]
\item
On
detection we switch the gates so that the particle cannot
jump back to the first island, and so on.
\\
\centerline{
\begin{picture}(60,10)(0,10)
\put(47,0){\line(0,1){10}}
\put(47,0){\line(1,0){10}}
\put(47,10){\line(1,0){10}}
\put(34,5){\circle{10}}
\put(12,5){\circle{10}}
\put(3,5){\circle*{1}}
\put(0,5){\circle*{1}}
\put(22,0){\usebox{\mysquare}}
\put(42,10){\usebox{\mysquare}}
\put(12,5){\circle*{5}}
\end{picture}
}\\
\end{enumerate}

 Since the steps (3)-(5) are
qualitatively similar to (1)-(3), we will only consider the first
step, that is, the transition from the state $0$ to the state $1$. During
the first process the energy of the particle was increased by
$E_1-E_0=V$ and this energy was taken from thermal fluctuations.  The
transitions between the right lead and the first island when the barrier is open are described by rates $\Gamma_{01}$ and $\Gamma_{10}$, 
satisfying the detailed balance relation $\Gamma_{10}/\Gamma_{01}=e^{V/T}$.

The measurement establishes when the particle jumps onto the first
island; it consists of a series of fast measurements at intervals
$\tau$. This can be realistic when there is a pulsed detector which is
sensitive to the presence or absence of a particle only during short
measurement periods. Alternatively, we can consider a continuously
working detector, but with a certain response time $\tau$. An ideally
continuous process corresponds to $\tau \to 0$.

\section{Thermodynamics 
and information flow}\label{sec:info}

\subsection{Extracted power, information  and dissipated heat}

After the particle is detected on the first island, 
the energy $E_1-E_0=V$ 
 is
available to do mechanical work. Therefore, the average power which
can be extracted, is the ratio $V/\Theta$ where $\Theta$ is the
average time per step. The probabilities
$p_0=1-p_1$ and $p_1$ to find the system in state 0 and 1,
respectively, satisfy the master equations
\begin{equation}\label{eq:master}
 \dot{p}_0 =  -\Gamma
 p_0+\Gamma_{10}, \quad 
\dot{p}_1 =  -\Gamma  p_1+\Gamma_{01}  
\end{equation}
where $\Gamma = \Gamma_{01}+\Gamma_{10}$. The solution of
Eq.~\eqref{eq:master} with initial conditions $p_0(0)=1$, $p_1(0)=0$,
gives the probability $p_\tau=p_1(\tau)$ for detecting the particle in
state 1 at time $\tau$. The probability of detecting the particle in
state 1 at measurement $n$ and not before is $s_n = (1-p_\tau)^{n-1}p_\tau$.
The average number of trials, $\langle N \rangle$, is then
\begin{equation} \label{N}
\langle N \rangle = \sum_{n=1}^\infty ns_n =\frac{1}{p_\tau} =
\frac{\Gamma}{\Gamma_{01}(1- e^{-\Gamma \tau})}, \ \  \Theta = \langle N \rangle \tau\, .
\end{equation}
How much information is obtained in this process?
Each measurement has two outcomes, either to give the same result as the
previous measurement (with probability $1-p_\tau$) or to change to the
opposite result (with probability $p_\tau$). These are independent for each
measurement, and therefore the information per measurement is
\begin{equation}\label{S1}
  S_1 = -p_\tau \ln p_\tau -(1-p_\tau)\ln (1-p_\tau) \, .
\end{equation}
We assume the measuring device to be error free. Otherwise the
information content should be characterized by the \textit{mutual}
entropy~\cite{Sagawa2010}.  Since $p_\tau=\langle N \rangle^{-1}$ and we
need $\langle N \rangle $ repeated measurements before a particle is
detected on the first island we get in the limit $\langle N \rangle
\gg1$ that $S=\langle N \rangle S_1\approx\ln \langle N \rangle
$. This is the number of binary digits needed to store $\langle
N\rangle$ times $\ln 2$, the information per bit.  Equation \eqref{N}
yields $S \to \infty$ for continuous measurement, $\tau \to 0$,
see also~\cite{Strasberg2013}. At $\tau \to \infty$, $S$
reaches its equilibrium value tending to $V/T$ at large $V$.

In order to return the system, including the measurement device to the
initial state, the information gained during measurement must be
deleted. According to Landauer's principle, this must lead to a
dissipation of heat to the environment. Using, Eqs. \eqref{N} and
\eqref{S1} and the fact that detailed balance gives
\[
 \frac{\Gamma_{01}}{\Gamma} = \frac{1}{1+e^{V/T}}
\] 
we get that the work required in deleting the information is 
\begin{equation}\label{heat}
\begin{aligned}
 W_{\text{delete}} &= TS = T\ln\left(1+e^{V/T}\right)
 -T\ln\left(1-e^{-\Gamma\tau}\right) \\
 &- T\frac{e^{
     V/T}+e^{-\Gamma\tau}}{1-e^{-\Gamma\tau}}\ln\frac{e^{
       V/T}+e^{-\Gamma\tau}}{1+e^{V/T}}\, .
\end{aligned}
\end{equation}
The arguments of the logarithms in the last two terms are both less
than 1, and the two logarithms are therefore both negative. The two
last terms give then  positive contributions, and  we have
\[
W_{\text{delete}} > T\ln\left(1+e^{V/T}\right) > V = \Delta U.
\] 
Here
$\Delta U$ is the change in internal energy of the system. This is
available for converting to mechanical work at the end of the process.
The work needed to delete the information is always greater than the
work we could extract using this information. Only in the limit
$\langle N\rangle\gg1$ (or $V \gg T$) and $\Gamma\tau\gg1$ do we get
that
\begin{equation}\label{heat2}
 W_{\text{delete}} =   T \ln\langle N\rangle   = T\ln\left(1+e^{\Delta
 E/T}\right)  \approx  V \, .
\end{equation}
Thus in the above limiting case the dissipated heat is the same as the
work, which can be extracted. In other cases is seems that our device
is operating not optimally. This would mean that some parts of the
process are irreversible and generate net entropy in the surroundings.

\subsection{Entropy production in irreversible expansion.}

Where does the entropy production occur?
Every time we measure and find the particle not on the first island we
know that it is on the right lead (including before the first
measurement). But we do not use this information to get energy, we
open the barrier (or keep it open). This is analogous to the free
expansion of a gas, which is an irreversible process, leading to net
increase of the total entropy. Indeed, at the beginning of
the cycle, or after a measurement which did not detect a transition to
state 1, the entropy is 0. Then we let the system evolve with the
barrier open, and the entropy will increase. After the time $\tau$,
when the next measurement is performed, it has the value given by
\eqref{S1}.  This is an irreversible process, implying that we could
arrive at the same final state in a different way, which would allow
us to extract some work $W_{\text{ex}}$ during the transition process,
\textit{in addition} to the increased internal energy $\Delta U = V$,
which is available after the transition took place. According to the
Landauer principle, if the information entropy in the end of the MD
operation is deleted to the same heat bath, the full available work
becomes non-positive.

In the limit of slow operation we can find the work $W_{\text{ex}}$
which could in principle be extracted during the transition process.
Imagine it somehow performed reversibly. That would mean to go through
some process that starts with the particle in the right lead (state 1)
and ends with the particle statistically distributed between the right
lead and the first island.

In this process, energy in the form of heat $Q=T\Delta S$ (where
$\Delta S$ is the change of entropy of the system) would be taken from
the thermal reservoir and work $W_1$ be extracted. In the limit of infinitely
slow operation the probability $p_1$ in the final
state is given by the thermal equilibrium values
\[
 p_1 = \frac{1}{e^{v}+1}
\]
where $v =  V/T$.

 The work that could be extracted is $W_1=T\Delta S - \Delta
U$ where 
\[
 \Delta U = \frac{ T v }{e^{v}+1}
\]
is the average increase in internal energy.
Substituting
\[
T\Delta S = -\frac{T }{e^{v}+1} \left( \ln
\frac{1}{e^{v}+1} +e^v \ln \frac{1}{e^{-v}+1}
\right)
\]
we obtain
\begin{equation} \label{work}
-\frac{W_1}{T}= \ln \frac{1}{e^{-v}+1} =  \ln (1-p_1) =  \ln
\left[1-\frac{1}{\langle
 N\rangle}\right] .
\end{equation}
Equation \eqref{work} gives the increase in entropy when opening the
barrier once or keeping it open when we know that the particle is
\textit{not} on the island. In total, therefore, the work which could
be extracted in a reversible transition is
\begin{equation}\label{waste}
 W_{\text{ex}} = -T\langle N\rangle\ln\left[1-\frac{1}{\langle
 N\rangle}\right] \, .
\end{equation}
The information that we got but did not use must still be deleted from
the memory, and the total work spent on deleting the memory is 
\begin{eqnarray} \label{del}
  W_{\text{delete}} &=& -T\langle N\rangle \left[p_\tau \ln p_\tau +(1-p_\tau)\ln(1-p_\tau)\right]
\nonumber \\ &=&
T\ln\langle N\rangle-T(\langle
  N\rangle-1)\ln\left[1-\frac{1}{\langle N\rangle}\right]  .
\end{eqnarray} 
Here we have taken into account that $p_\tau=1/\langle N \rangle$.
Combining Eqs.~\eqref{waste} and \eqref{del} 
we can express the difference between the work in
deleting and the extracted work as 
\[
 W_{\text{delete}}  -(  W_{\text{ex}} +\Delta U) =  T \ln(\langle N\rangle - 1)-V\, .
\]
Here we recall that the extracted energy $\Delta U= V$ is
equal to the energy in the final state when the particle is detected
on the first island.
Using Eq.~\eqref{N} we get
\begin{equation}\label{energyBalance}
 W_{\text{delete}}-(  W_{\text{ex}} +\Delta U)  = 
 T\ln\frac{1+e^{-\Gamma\tau}e^{-V/T}}{1-e^{-\Gamma\tau}} \, .
\end{equation}
The above expression
 shows explicitly that when $\Gamma\tau\gg1$ we get $ W_{\text{delete}}  = W_{\text{ex}} +  \Delta U$.

The limit $\Gamma\tau\gg1$ corresponds to slow operation of the
device, so it is similar to the usual requirement of quasi-static
operation for reversible processes. It should be noted that the extracted
work \eqref{waste} was calculated in the limit of quasistatic
operation, so that for finite operation time, $\tau$, we would have a
smaller amount of work that could be extracted because some entropy
would be created. This means that in Eq.~\eqref{energyBalance},
$W_{\text{ex}}$, which is the work that is wasted because of our
protocol, should be less, since at finite operation rate also any
other protocol would be suboptimal. What is the maximal amount of work
which can be extracted in a finite time? Or equivalently, what is the
minimal entropy production? This question will be answered in the
following section.

\section{Optimal protocol}\label{sec:optimal}

We implement the idea of reversible expansion of the configuration space \cite{horowitz2011} in the
following way.  (a)~We start form the configuration when the particle
is on the right lead and the barrier is closed. (b)~Then we quickly
lift the potential of the first island to a high value $V_0\gg T$. (c)~The potential of the left side is slowly moved
down. At this stage the barrier gradually opens and transitions between the states
can take place. (d)~Lowering of the potential continues until time $\tau$ when the next measurement is due. At this time,
some energy $V(\tau)$ is reached. (e)~If the measurement shows absence
of the particle, the voltage is quickly increased again, and the
process is repeated.  If the particle has been detected,
a similar process is started at the adjacent grain.
The raise of the potential at stage (b) occurs faster than the transition time $\Gamma^{-1}$ at stages (c) and (d) but slower than the relaxation in the heat bath, which is the fastest time in our system. Lowering of the barrier (or the measurement time $\tau$) is slower than $\Gamma^{-1}$.

Let us denote $E_i(t)$ the energy of state $i$ as
function of time, The protocol described above has $E_0(t)=0$ and
$E_1(t)=V(t)$. Let $p_i(t)$ be the probabilities to find the particle
in state $i$. These satisfy the master equation~\eqref{eq:master} with the rates $\Gamma_{ij}$ now depending on
the difference $V(t)=E_1(t)-E_0(t)$ and, therefore, on time. The extracted work during time $\tau$ is
\[
 W_{\text{ex}} = -\sum_i \int_0^\tau dt \,  p_i(t)\dot{E}_i \, .
\]
The average internal energy change is 
\[
 \Delta U = \sum_i[p_i(\tau)E_i(\tau)-p_i(0)E_i(0)] \, .
\]
The heat transfer is then 
\begin{equation}\label{Q}
 Q = \Delta U + W_{\text{ex}} = \sum_i \int_0^\tau dt \, \dot{p}_i E_i(t) \, .
\end{equation}
Since the thermal bath is never brought
out of equilibrium due to fast relaxation, the change in entropy of the
environment is $\Delta S_{\text{env}} = -Q/T$. The entropy of the
system is $ S = -\sum_i p_i\ln p_i$.
As in~\cite{Aurell2011,aurell2012} we 
write the change in entropy as an integral
\begin{equation}\label{eq:entropy}
 \Delta S = -\sum_i \int_0^\tau dt \, \frac{dp_i}{dt}\ln p_i\, .
\end{equation}
This is the information entropy stored in the measuring device. It
has to be deleted in the end to reset the device.  Being interested
here in optimizing the losses in course of extracting the work, we do
not consider losses in the process of deleting the information, which should
also be done in an optimal way.  In general,
Eq.~\eqref{eq:entropy} defines $\Delta S$
as a \textit{functional} of the operation protocol $E_i(t)$. 
We need also to solve the master
equation~\eqref{eq:master}. This is complicated by the fact the
transition rates $\Gamma_{ij}$ depend on the energy difference $V(t)$
between the two sites maintaining detailed balance.  By specifying the
time dependence $V(t)$ we then get time-dependent $\Gamma_{ij}(t)$. The
sum of the rates is $\Gamma(t)
=\left(1+e^{V(t)/T}\right)\Gamma_{01}(t)$. 
The master equation 
\begin{equation}\label{eq:me}
 \frac{dp_1}{dt} = -\Gamma(t) p_1+\Gamma_{01}(t) 
\end{equation}
can now be integrated for any known dependence $V(t)$. With the
initial condition $p_1(0)=0$ we get
\begin{equation}\label{eq:masterSol}
 p_1(t) = \int_0^tdt'' e^{\int_t^{t''}dt'\Gamma(t')}\Gamma_{01}(t'')\, .
\end{equation}
As in~\cite{Mandal2012},
to simplify the following calculations we choose $\Gamma$ to be
independent of $V$: $$\Gamma \equiv \gamma_0\, \quad  
\Gamma_{01}(t)=\gamma_0
\left(1+e^{V(t)/T}\right)^{-1}$$ which does not affect the results qualitatively.

\subsection{Optimized entropy production for fixed heat flux.}\label{protocol}

We now return to Eq.~\eqref{eq:entropy} and want to optimize it in the
sense of finding the operation time $\tau$ and protocol $V(t)$, which
will minimize the entropy production. It is clear that if this is done
without any constraints, the entropy production can be made arbitrary
small by choosing $\tau$ and $V_0$ large enough. This would mean that
we are in the quasistatic regime discussed above. In this case the
produced power is zero since we get a finite amount of energy in an
infinite time. A more instructive situation would be to minimize the
entropy production rate at a constant heat flux. If we imagine the Maxwell
demon device as part of a heat engine, with the reservoir supplying
the energy to the particle as a high temperature reservoir and the
reservoir at which we delete the memory of the demon as a low
temperature reservoir, this would correspond to maximizing the output
power at a given input heat.

\subsubsection{Derivation of the entropy production functional and the
integral equation for the optimal protocol}

The power is now defined as the average heat per cycle extracted from
the reservoir, $\mathcal{P}=Q/\tau$. The total entropy is $\Delta
S_{\text{tot}} =\Delta S+\Delta S_{\text{env}}$ and the rate of
entropy production is then
\begin{equation}\label{eq:entProd}
\frac{\Delta S_{\text{tot}}}{u_0} = \frac{\Delta S}{u_0}-
\frac{Q}{u_0 T}
  = \frac{\Delta S}{u_0}- P_0\, .
\end{equation}
where we introduce the dimensionless variables
\begin{equation}\label{eq:dimLessVar}
 u = \gamma_0 t, \ \ u_0=\gamma_0\tau, \  \ v (u)= V(t)/T, \ \
P_0 = \mathcal{P}/\gamma_0T.
\end{equation}
Since $P_0$ is to be held constant, it is
sufficient to minimize $ \Delta S/u_0$. 
Denoting $p \equiv p_1$ we obtain:
\begin{eqnarray*}
 \frac{\Delta S}{u_0} &=& -\frac{1}{u_0}\int_0^{u_0}du\left[ \ln p\, 
 \frac{dp}{du} + \ln(1-p)\, \frac{d(1-p)}{du}\right] \\
&=& -\frac{1}{u_0}\int_0^{u_0}du \, \ln \left(\frac{p[v]}{1-p[v]}\right)
\frac{dp[v]}{du}
\end{eqnarray*}
where from \eqref{eq:masterSol}
\begin{equation}\label{eq:p}
 p [v]= \int_0^u du' \, \frac{e^{u'-u}}{e^{v(u')}+1}\, .
\end{equation}
This should be minimized subject to the constraint that the power $P_0$
is a given constant. Using Eq. \eqref{Q} and the master equation~\eqref{eq:me} 
we find that it can be expressed as
\begin{equation}\label{constr}
P_0 = \frac{1}{u_0}\int_0^{u_0}du\left(-p[v]
+\frac{1}{e^{v(u)}+1}\right)v(u)\, .
\end{equation}
Introducing the Lagrange multiplier $\lambda$ this means that we have
to minimize the functional 
\begin{equation} \label{Ifunctional}
I = \frac{1}{u_0}\int_0^{u_0}du\left(\lambda v - \ln
   \frac{p}{1-p} \right)   
   \left(-p+\frac{1}{e^v+1}\right)
\end{equation}
with the function $v(u)$ and the 
 time $u_0$ as variables. 

The functional $I$ is of the form
\begin{equation}\label{action}
 I[v] = \frac{1}{u_0}\int_0^{u_0}du \, L(v,p[v])
\end{equation}
with the function
\begin{equation}\label{lagrange}
 L(v,p[v]) = \left(\lambda v - \ln
   \frac{p[v]}{1-p[v]} \right) 
   \left(-p[v]+\frac{1}{e^{v}+1}\right)
\end{equation}
depending on $v$ both directly and
indirectly through $p[v]$ which is a functional of $v$ as given by
Eq.~\eqref{eq:p}. The Euler-Lagrange equation takes the form
\begin{equation}\label{eq:inteq}
 \left.\frac{\partial L}{\partial
v}\right|_u - \int_u^{u_0}du'\frac{e^{u'-u}}{(e^{v'}+1)^2}e^{v'}\left.\frac{\partial L}{\partial
p}\right|_{u'} = 0\, .
\end{equation}
Calculating the derivatives, 
 we get the Euler-Lagrange
equation as
\begin{equation}\label{eq:inteq2}
\begin{aligned}
 &\lambda\left[v+\left(p-\frac{1}{e^v+1}\right)\frac{(e^v+1)^2}{e^{v}}
    - e^u\int_u^{u_0} \! \! \!  du'\, e^{-u'}v'\right]\\
  &= \ln\frac{p}{1-p}-e^u
  \int_u^{u_0} \! \! \! du'e^{-u'}\left[\frac{1}{p'(1-p')}
    \left(p'-\frac{1}{e^{v'}+1}\right)  \right.\\
& \left. \quad +  \ln\frac{p'}{1-p'}\right]
\end{aligned}
\end{equation}
where $p'= p[v(u')]$. This equation is probably impossible to solve
analytically and to proceed we consider operation of the device which
is sufficiently slow so that the probability $p$ never is far from
the thermal equilibrium value.

\subsubsection{Lowest order correction to the quasistatic solution}

If we consider slow operation the deviation from the quasistatic
solution will be small. That is, we write $p=p_a+p_b$ where
%
\[
 p_a = \frac{1}{e^v+1}
\]
and the Master equation \eqref{eq:me} is 
%
$ \dot p = -p+p_a.$
For later use it is convenient to introduce the more general Master
equation 
\begin{equation}\label{eq:meGeneral}
 \dot p = -g(v)(p-p_a)
\end{equation}
where $g(v)$ is some function of $v$ and the present case corresponds to $g(v)=1$. 
Neglecting $\dot p_b$ compared to $\dot p_a$ we get
\begin{equation}\label{pb}
 p_b = -\frac{1}{g(v)}\, \dot{p}_a =  -\frac{1}{g(v)}\frac{dp}{dv}\, \dot{v}
 = \frac{\dot{v}e^v}{g(v)\, (e^v+1)^2}.
\end{equation}
Inserting this into the Lagrangian \eqref{lagrange} we get 
\begin{equation} \label{lag1}
  L(v,\dot v) =  \alpha(v)\dot v^2+ \beta(v)\dot v
\end{equation}
with 
\begin{eqnarray} 
\alpha (v) &=& \frac{e^v}{g(v) (e^v+1)^2} = \frac{1}{4g(v) \cosh^2v/2},  \label{alpha} \\
\beta (v ) & =& -\frac{\lambda+1}{g(v)}\frac{e^v}{(e^v+1)^2}  
   =  -(\lambda+1)vg(v)\alpha(v) \nonumber \\
&=&  -(\lambda+1)\frac{v}{\dot{v}}\, p_b (v). \label{beta}
\end{eqnarray}
The Euler-Lagrange equation~\eqref{eq:inteq2} can be  integrated to give 
\begin{equation}\label{eq:vdot}
 \dot{v} = \frac{A}{\sqrt{\alpha(v)}}
\end{equation}
where $A$ is a constant of integration. A second integration gives 

\begin{equation}\label{F}
 F(v) = \int_{v_0}^v dv\sqrt{\alpha(v)}= Au
\end{equation}
where $v_0=v(0)$ which at the moment is unspecified. 

Since the values of $v$ at the  endpoints are not fixed we get the
additional conditions 
\[
 \left. \frac{\partial L}{\partial \dot v}\right|_0 = 
 \left. \frac{\partial L}{\partial \dot v}\right|_{u_0} = 0\, .
\]
Since
\[
 \partial L/\partial \dot v = 2\alpha(v)\dot v+\beta(v)
 = \alpha(v) [2\dot v-(\lambda+1)vg(v)]=0
\]
the equality $ \partial L/\partial \dot v =0 $ can be met either at
$\alpha(v)=0$ or at $\dot v =
\frac{1}{2}(\lambda+1)vg(v)$. $\alpha(v)=0$ implies that $v=\infty$ and we
guess that this is the proper solution at $u=0$, so that we have $v_0
\equiv v(0)=\infty$. 
At the final time, $u_0$, we assume that $v_u= v(u_0)$ is finite which
means that we must have
\[
 \dot{v}(u_0) = \frac{1}{2}(\lambda+1)v_ug_u
\]
where $v_u = v(u_0)$ and $g_u = g(v_u)$. Using \eqref{eq:vdot} we get 

\begin{equation}\label{eq:Alambda}
 A  = \frac{1}{2}(\lambda+1)v_ug_u\sqrt{\alpha} =
 \frac{v_u(\lambda+1)}{4\cosh v_u/2}\, .
\end{equation}
The equation \eqref{constr} for the constraint takes the form 
\begin{equation}\label{p0}
 u_0P_0 = -\int_0^{u_0} \! \! \! du \, p_bv = \int_{v_u}^\infty \! \! \!
 \frac{dv \, ve^v}{g(v) (e^v+1)^2} = K(v_u).
\end{equation}
Finally, the action \eqref{action} should also be stationary with
respect to
variation of the operation time $u_0$. The condition $\partial
I/\partial u_0 =0$
can be rewritten as

\begin{equation}\label{u0}
  L(v_u,\dot v_u) = \frac{1}{u_0}\int_0^{u_0}duL(v,\dot v) \, .
\end{equation}
Using \eqref{eq:vdot} we find
\[
 L = \alpha(v)\dot{v}^2+\beta(v)\dot{v} = A^2 + \dot{v} \, .
\]
Using \eqref{beta} and \eqref{p0} we find 
\[
 \frac{1}{u_0}\int_0^{u_0} du \beta(v)\dot{v} = (\lambda + 1)P_0\, .
\]
Equation \eqref{u0} is then 
\[
 -v_u p_b(v_u)  = P_0\, .
\]
Using \eqref{pb}, \eqref{eq:vdot} and \eqref{p0} we get 
\[
 Au_0 = -\frac{2K(v_u)}{v_u}\cosh \frac{v_u}{2}\, .
\]
From \eqref{F} we have $F(v_u) = Au_0$, and we get an equation in terms
of $v_u$ only 
\begin{equation}\label{eq:vu}
 F(v_u) = -\frac{2K(v_u)}{v_u}\cosh \frac{v_u}{2} \, .
\end{equation}
\begin{figure}[b]
\centerline{
\includegraphics[width=.6\linewidth]{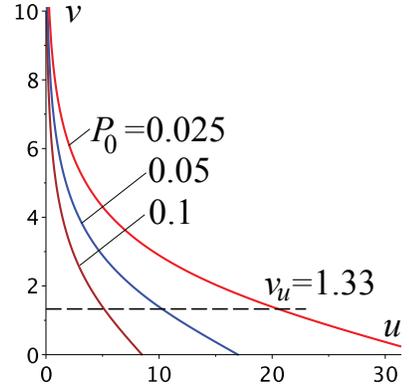}
}
\caption{(Color online) Plot of the optimal protocol $v(u)$ for different output power
$P_0$. The optimal period, $u_0 \equiv \gamma_0\tau$, between
measurements is $\approx 0.51/P_0$,
the final value of $v$ is $\approx 1.33$.
 \label{fig1}}
\end{figure}
This equation must in general be solved numerically to give $v_u$. It
is interesting to note that $P_0$ does not enter the equation, which
means that $v_u$ does not depend on $P_0$. 
 All other quantities are expressed in terms of the
solution $v_u$:
%
$$ u_0 = \frac{K(v_u)}{P_0}, \ 
 A = -\frac{2P_0}{v_u}\cosh \frac{v_u}{2}, \
 \lambda = -\frac{8P_0}{v_u^2}\cosh^2 \frac{v_u}{2} -1 .$$
We can also calculate the rate of entropy generation as a function of
the power $P_0$. We have Eq. \eqref{eq:entProd} and using \eqref{lag1}
with $\lambda = 0$ we find that
\[
\frac{\Delta S}{u_0} = \frac{1}{u_0}\int_0^{u_0}du
\left.L(v,\dot{v})\right|_{\lambda=0}  = A^2 + P_0\, .
\]
Therefore 
\[
\frac{\Delta S_\text{tot}}{u_0} = A^2 =
\frac{4\cosh^2(v_u/2)}{v_u^2}P_0^2\, .
\]
The rate of entropy production is quadratic in $P_0$. 
In the case $g(v) = 1$ we get
\[
 F(v) = - \int_v^\infty dv \frac{e^{v/2}}{e^v+1} =
 - \arccot(\sinh\frac{v}{2})
\]  

and 
\begin{equation}\label{Kvu}
 K(v_u)  =  \int_{v_u}^\infty \! 
 \frac{dv\,  ve^v}{(e^v+1)^2} =
 \ln\left(e^{v_u}+1\right)-\frac{v_u}{1+e^{-v_u}} .
\end{equation}
Equation~\eqref{eq:vu} 
then
becomes
%
$$
 \arccot\left(\sinh\frac{v_u}{2}\right ) =
\frac{\cosh ( v_u/2)}{v_u/2} 
\left[\ln \left(e^{v_u}+1\right) 
  - \frac{v_u}{1+e^{-v_u}}\right]\! .
$$
This equation can be solved numerically giving $v_u =  1.3256$.
From this we get  $K=0.5138$ and 
\begin{equation}\label{lambda}
\lambda = -6.8630P_0-1 \qquad \mbox{and}  \qquad  A =  -1.8524 P_0 \, .
\end{equation}
The optimal protocol is plotted in Fig.~\ref{fig1} (left) for various values of $P_0$.

The entropy production per measurement interval is $\Delta S_{\text{tot}}/\tau =3.43 \gamma_0 P_0^2$.
The extracted work is 
$
\Delta W_{\text{ex}}/\tau =0.46 {\cal P}
$ neglecting the quadratic term. The average operational time $\Theta = 2.44 T/{\cal P}$.
Our results hold for slow processes, $u_0\gg 1$ or ${\cal P}\ll T\gamma_0$.

\subsection{Leads with many states and particles}

\begin{figure}[b]
\setlength{\unitlength}{1mm}
\begin{center}
\includegraphics[width=0.8\linewidth]{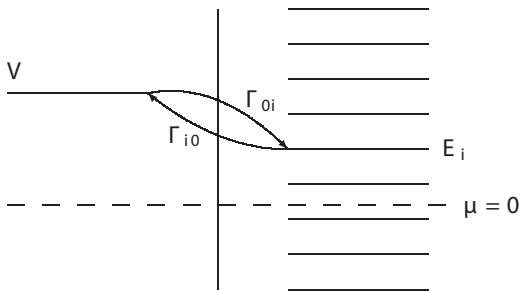}
\end{center}
\caption{The more realistic model with a single level on the first
  island (left) and a band with many levels in the metallic lead
  (right). The Fermi level is chosen as $\mu=0$ and the rates for
  transitions between the island and level $i$ are $\Gamma_{0i}$ and
  $\Gamma_{i0}$.\label{fig:manyLevels}}
\end{figure}

Let us now consider a more realistic model where the leads have a band
of states with different energies, and we have many particles in the
lead as illustrated in Fig.~\ref{fig:manyLevels}. 
We make the following assumptions: 
\begin{enumerate}
\item{Detailed balance: 
$
 \Gamma_{0i} = \Gamma_{i0}e^{(V-E_i)/T}.
$
}
\item{$ \Gamma_{0i} + \Gamma_{i0}=\Gamma_0$ where $\Gamma_0$ is a
    constant independent of $V$ and $E_i$. Together with the detailed
    balance this implies
\begin{equation}
 \Gamma_{i0} = \frac{\Gamma_0}{1+e^{(V-E_i)/T}}  \, , \quad  
 \Gamma_{0i} = \frac{\Gamma_0}{1+e^{-(V-E_i)/T}}.
\end{equation}}
\item{Internal processes in the lead are fast so that the lead is
    always in equilibrium, which means that the probability of finding
  the level at $E_i$ occupied is 
\begin{equation}
 f(E_i) = \frac{1}{e^{E_i/T}+1}\,  .
\end{equation}
The chemical potential $\mu=0$.}
\item{The density of states in the lead is constant: $g(E_i) = g_0$}
\end{enumerate}
Under these assumptions, the entropy of the leads is constant. If $p$
denotes the probability that an electron is found on the island, the
change in entropy is $\Delta S = -p_\tau\ln p_\tau -(1-p)_\tau \ln (1-p_\tau)$.  The master equation
is
\begin{equation}
\dot{p} = \underbrace{\sum_i\Gamma_{i0}f(E_i)}_{\Gamma_{12}(V)}(1-p) - \underbrace{ \sum_i\Gamma_{0i}(1-f(E_i))}_{\Gamma_{21}(V)}p
\end{equation}
where
%
\begin{eqnarray}
\frac{\Gamma_{12}(V)}{\Gamma_0} &=&
\int_{-\infty}^\infty \! 
    \frac{g_0\, dE}{(1+e^{(V-E)/T}) (e^{ E/T}+1)}
= \frac{g_0V}{e^{V/T}-1}, \nonumber \\
\Gamma_{21}(V) &=& -\frac{\Gamma_0g_0V}{e^{- V/T}-1} =
\Gamma_{12}(V)e^{V/T}\, .
\end{eqnarray}
Denoting $ \gamma_0 = 2T\Gamma_0g_0$ and introducing  dimensionless variables as in \eqref{eq:dimLessVar} we get
$$\Gamma =  \Gamma_{12}+ \Gamma_{21} = \gamma_0\frac{v}{2}\coth\frac{v}{2}\ \text{and} \   \Gamma_{12} = \frac{\Gamma}{1+e^v}\, .$$
This is still a model. For some microscopic mechanisms $\Gamma \propto
E^p$ where $p$ is some number. Then some extra power of $v$ can appear
in the expression for $\Gamma$.

In the limit of slow operation we find that the Master equation
\eqref{eq:me} is
\begin{equation}
 \dot{p} = \frac{dp}{du}  = -\frac{v}{2}\coth\frac{v}{2}(p-p_a)
\end{equation}
which is of 
the general form \eqref{eq:meGeneral} with 
\begin{equation}
 g(v) = \frac{v}{2}\coth\frac{v}{2}\, .
\end{equation}
This means that the more realistic model with the leads modeled as
metallic bands has the same features as the simplified model discussed
in Sec.~\ref{protocol}.
According to Eq. \eqref{F} we get
\begin{equation}
 F(v) = -2\int_v^\infty dv\frac{e^{v/2}(e^v-1)}{v(e^v+1)^2}
\end{equation}
which cannot be solved in closed form. 
From Eq.~\eqref{p0} we have 
\begin{equation}
 K(v_u) = \frac{1}{2\cosh^2v_u/2}
\end{equation}
and Eq.~\eqref{eq:vu} gives $v_u = 1.5076$.


\subsection{Optimal protocol for a fixed energy gain.}

\begin{figure}[t]
\begin{center}
\includegraphics[width=.6\linewidth]{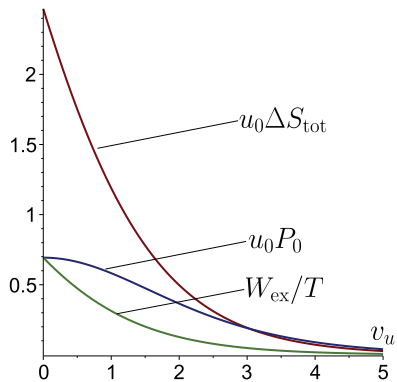}
\end{center}
\caption{(Color online)  Normalized heat flux $u_0P_0$, the entropy production 
$u_0\Delta S_{\text{tot}}$, and extracted work $W_{\text{ex}}/T$
as functions of the given energy gain.
 \label{fig3}}
\end{figure}

To raise the system up by a given energy $V$ using one island, we need
to minimize the entropy production keeping the final energy fixed,
$v_u =V/T$.  For a given time between the measurements, $\tau$, this
will determine the required heat flux and the entropy production. As
can be seen from Eqs.~\eqref{eq:entProd},~\eqref{action}, the total
entropy production in this case is determined by the functional $I[v]$
taken at $\lambda =-1$.  The solution of the Euler-Lagrange equation
is again given by Eq.~(\ref{F}) [see Fig.~\ref{fig1}]. The
entropy production and the required heat flux are easily calculated in
the same way as before: $P_0 = K(v_u)/u_0$, $\Delta
S_{\text{tot}}/u_0 = [F(v_u)]^2/ u_0^2$ where $F(v_u)$ and
$K(v_u)$ are determined by Eqs.~(\ref{F}) and (\ref{Kvu}), see
Fig.~\ref{fig3}.
  The measurement interval has to be longer
than $\gamma_0^{-1}$, i.e., $u_0 \gg 1$. For $v_u
\lesssim 1$, the heat flux is ${\cal P}\ll T\gamma_0$. The average
operational time needed to complete the transition, $\Theta = \tau /p(v_u)=
\tau (e^{v_u} +1)$, grows exponentially for high energy gain
$V\gg T$.  To optimize the total operational time, it is thus
favorable to divide the total energy interval $\Delta E_{\text{fin}}$
into $M$ steps such that each step has the height $V\sim
T$. The device thus will have $M-1$ island with the total operational
time proportional to $M$ instead of being exponential. Minimizing $\Theta= \tau M\left(e^{\Delta E_{\text{fin}}/MT}+1\right)$ we
find that the optimal number of steps is the closet integer to $M =1.28
\Delta E_{\text{fin}}/T$.  The extracted work is
$$W_{\text{ex}}/\tau = (T/\tau) [K(v_u)-v_u p(v_u)]; $$
the ratio $W_{\text{ex}}/T $ is shown in
Fig.~\ref{fig3}.  Since $K(0) =\ln 2$ we find that, for zero
energy gain $v_u \to 0$, the extracted work assumes the value
$W_{\text{ex}} = {\cal P}\tau = T\ln 2$ as for the symmetric
Szil{\'a}rd engine.

\section{Discussion and conclusion}\label{sec:disc}

This simplified model of a MD device allows us to make several general
conclusions applicable to a wider range of situations.  It is clear
that detection, feedback, and erasure of information should all be
optimized.  The position of the particle is a degree of freedom with
thermal fluctuations. The detector registers these fluctuations and
transfers the information into some
non-fluctuating degree of freedom (or fluctuating on a very long time
scale).  We observe that the rate $1/\tau$ at which such a degree of
freedom reads out the fluctuating quantity should not be larger than
the rate $\Gamma$ of transitions between the states of the fluctuating
degree of freedom. If $\Gamma\tau\lesssim1$ the readout does not
produce sufficient new information, while this information still has
to be deleted. We have also seen that an optimal protocol is needed to
minimize the irreversible entropy production in the course of opening
the gates between the parts of the device, thereby enlarging the
available configuration space. We construct such protocols for a given
input power or a given energy gain. Our conclusions are relevant to
many different devices, unrelated to the MD, where an optimized
protocol is needed to reduce the production of entropy, i.e., 
the dissipated heat.  The entropy production
rate at small heat flux is proportional to the heat flux
squared. Therefore, the ratio of the entropy production to the heat
flux vanishes at small heat fluxes when the device operates
reversibly.

\begin{acknowledgments}
We thank D.~V.~Averin,  J.~P.~Pekola, and I.~M.~Khaymovich for fruitful discussions. The research leading to these results has received funding from the European Union
Seventh Framework Program (FP7/2007-2013) under grant agreements No. 308850
(INFERNOS) and 228464 (MICROKELVIN).
\end{acknowledgments}

\end{document}